\documentclass[
  aps,prl,twocolumn,
  nofootinbib,nobibnotes
]{revtex4-1}

\usepackage[utf8]{inputenc}
\usepackage{amssymb}
\usepackage{amsmath}
\usepackage{amsfonts}
\usepackage{graphicx}
\usepackage{color}
\usepackage{xspace}
\usepackage{comment}
\usepackage{hyperref}
\usepackage[normalem]{ulem}

\usepackage[section]{placeins}
\usepackage{afterpage}

\usepackage{float}
\usepackage{slashed}
\usepackage{ulem}
\usepackage{appendix}

\usepackage{cancel}

\usepackage{multirow,rotating}
\usepackage[dvipsnames]{xcolor}
\usepackage{orcidlink}   

\begin{document}
	
      \title{Same-sign dimuon probe of charged lepton flavor violation at electron–photon colliders}
      
      \author{Zhong Zhang\,\orcidlink{0009-0008-6848-9224}}
      \affiliation{School of Physics, Hefei University of Technology, Hefei 230601, People’s Republic of China}

      \author{Yu Zhang\,\orcidlink{0000-0001-9415-8252}}
      \email{dayu@hfut.edu.cn}
      \affiliation{School of Physics, Hefei University of Technology, Hefei 230601, People’s Republic of China}
 
	   \author{Zeren Simon Wang\,\orcidlink{0000-0002-1483-6314}}
	   \email{wzs@hfut.edu.cn}
      \affiliation{School of Physics, Hefei University of Technology, Hefei 230601, People’s Republic of China}

	\begin{abstract}
        Observation of charged lepton flavor violation would constitute unambiguous evidence for physics beyond the Standard Model (SM). We identify a previously unexplored same-sign dimuon signature in electron--photon collisions, $\gamma e^- \to e^+\mu^-\mu^-$, mediated by an axionlike particle (ALP) with flavor-violating $e$--$\mu$ couplings. The absence of irreducible SM backgrounds and the on-shell production of the ALP render this channel intrinsically clean and highly sensitive, with only small residual backgrounds arising from detector effects. Such collisions can be realized via laser Compton backscattering at $e^+e^-$ colliders including BEPC-II with the BESIII detector, STCF, CEPC, and ILC. We find that STCF, CEPC, and ILC can probe couplings one to two orders of magnitude below existing bounds. This combination of resonant production, vanishing irreducible background, and same-sign topology would be difficult to achieve in conventional $e^+e^-$ or hadron-collider environments, establishing electron--photon collisions as a uniquely powerful probe of charged lepton flavor violation.
 	\end{abstract}
 	\keywords{}
		

	\maketitle
    \noindent

\emph{\textbf{Introduction}---}
Charged lepton flavor violation (cLFV) is highly suppressed in the Standard Model (SM), and its observation would constitute unambiguous evidence for new physics.
In this work, we identify a previously unexplored same-sign dimuon topology in electron--photon collisions, $\gamma e^- \to e^+\mu^-\mu^-$.
This process is mediated by an axionlike particle (ALP) with flavor-violating $e$--$\mu$ couplings, which can be produced on shell, leading to a resonant enhancement of the signal rate.
Moreover, the SM does not yield this final state at the parton level, rendering the signature intrinsically clean, with only small residual backgrounds arising from detector effects.
We find that this channel can probe cLFV ALP couplings one to two orders of magnitude beyond existing constraints at future facilities.
This work establishes a new collider paradigm for probing charged lepton flavor violation using electron--photon collisions, providing a uniquely powerful probe of cLFV.
This combination of resonant production already at leading order, vanishing irreducible backgrounds, and a distinctive same-sign topology would be difficult to realize in conventional $e^+e^-$ or hadron-collider environments.

Electron--photon collisions can be realized via laser Compton backscattering (LCB)~\cite{Arutyunian:1963xib,Milburn:1962jv,Ginzburg:1982bs,Ginzburg:1981vm,Telnov:1989sd} at electron--positron colliders including the Beijing Electron--Positron Collider II (BEPC-II) with the Beijing Spectrometer~III (BESIII) detector~\cite{BESIII:2009fln}, the Super Tau--Charm Facility (STCF)~\cite{Achasov:2023gey,Ai:2025xop}, the Circular Electron Positron Collider (CEPC)~\cite{CEPCStudyGroup:2018ghi,CEPCStudyGroup:2023quu,CEPCStudyGroup:2025kmw,Ai:2025cpj}, and the International Linear Collider (ILC)~\cite{ILCInternationalDevelopmentTeam:2022izu}.
In this setup, a laser beam is scattered off the positron beam to generate a high-energy photon beam that subsequently collides with the incoming electron beam.
This technique has been employed for precision beam-energy measurements at BESIII~\cite{Achasov:2008gq,ZhangJianYong:2015vpa,Zhang:2016BESIIIBeamEnergy,Zhang:2017luo} and can be implemented at future facilities such as STCF~\cite{Achasov:2023gey}, CEPC~\cite{Tang:2020gmv,Chen:2025bie}, and ILC~\cite{Muchnoi:2008bx}.
While $\gamma e$ collisions have been explored in various contexts~\cite{Eboli:1993wg,Atag:2003wm,Benbrik:2024hsf}, including cLFV production of a CP-even neutral scalar mediator in Ref.~\cite{BhupalDev:2018vpr}, the specific same-sign dimuon topology $\gamma e^-\to e^+\mu^-\mu^-$ and its dedicated collider phenomenology have not been investigated.
In this context, compared with conventional $e^+e^-$ collisions, the $\gamma e$ mode offers several distinctive advantages for this channel: the mediator can be produced on shell already in the leading $2\to2$ subprocess, resulting in a resonant enhancement of the signal rate.
The asymmetric initial-state kinematics further enhances the sensitivity to light mediators produced near threshold while maintaining sufficient boost for their decay products to be experimentally resolved.
These features, together with the absence of irreducible SM backgrounds, lead to an exceptionally clean and sensitive probe of weakly coupled new physics.

The QCD axion, arising as pseudo--Nambu--Goldstone bosons of spontaneously broken global symmetries, are well-motivated candidates for new physics.
More generally, the ALPs appear in various extensions of the SM; unlike the QCD axion, their masses and couplings are largely independent parameters.
Their interactions need not respect the flavor structure of the SM, allowing for cLFV couplings that have been explored in various works; see, for example, Refs.~\cite{Cordero-Cid:2005vca,Heeck:2016xwg,Bauer:2019gfk,Cornella:2019uxs,Endo:2020mev,Calibbi:2020jvd,Cheung:2021mol,Araki:2022xqp,Calibbi:2024rcm,Ardu:2026vsr}.
In this Letter, we focus on ALPs with flavor-violating $e$--$\mu$ couplings and study the process $\gamma e^- \to e^+\mu^-\mu^-$ at electron--photon colliders.
We show that this same-sign dimuon channel provides a uniquely sensitive probe of cLFV, with sensitivities at STCF, CEPC, and ILC reaching one to two orders of magnitude beyond existing bounds.
Moreover, this sensitivity substantially surpasses that of projected searches at conventional $e^+e^-$ colliders such as Belle II, despite their much larger integrated luminosities.
This establishes electron--photon collisions as a uniquely powerful and previously unexplored probe of ALP-mediated charged lepton flavor violation.

\vspace{0.2cm}
\emph{\textbf{Theoretical setup}---}
We consider a real pseudoscalar ALP $a$ with flavor-violating couplings to electrons and muons, described within a model-independent low-energy effective framework.
The relevant interaction is given by
\begin{equation}
\mathcal{L}_{\rm int} = - i g_{ae\mu}\, a\, \bar{e}\gamma_5 \mu + {\rm h.c.},\label{eqn:lagrangian}
\end{equation}
where $g_{ae\mu}$ is a dimensionless coupling.
Here, we have considered a pseudoscalar interaction of the ALP.
In principle, a scalar-type operator can be studied instead; however, doing so would not qualitatively modify the collider sensitivity, since the two couplings lead to very similar kinematic structures and signal-event yields for the processes under consideration.

The off-diagonal couplings arise naturally in ultraviolet completions with flavor-dependent interactions or fermion mixing (see, e.g., Ref.~\cite{Calibbi:2020jvd}).

In general, ALP couplings are formulated in a derivative form.
However, upon integration by parts and using the fermion equations of motion, they can be recast into the Yukawa-like form in Eq.~\eqref{eqn:lagrangian} for on-shell external fermions, which suffices for the processes considered in this work.

We further note that, in generic ultraviolet completions, charged-lepton-flavor-conserving (cLFC) couplings may coexist with cLFV interactions, inducing the tightly constrained decay $\mu\to e \gamma$ at one loop.
Using the framework of Ref.~\cite{Calibbi:2024rcm} together with the latest MEG II bound~\cite{MEGII:2025gzr}, we estimate the corresponding limits on the diagonal ALP couplings.
We find that the resulting constraints are strongly dependent on the ALP mass and on the assumed flavor structure.
Representative limits on the flavor-conserving couplings are presented in the Supplemental Material.
In this work we take the standard single-coupling-dominance assumption and set the cLFC couplings to zero.

The decay width of the ALP into a charged-lepton pair is given by
\begin{equation}
\Gamma(a \to e^\pm \mu^\mp) = \frac{g_{ae\mu}^2}{8\pi} m_a \left(1 - \frac{m_\mu^2}{m_a^2}\right)^2.
\end{equation}
The total decay width is twice this value.
The corresponding proper decay length $c\tau_a$ depends on both the coupling strength and the ALP mass.
In the laboratory frame, the decay length is further scaled by the Lorentz boost factor, thereby determining the relative importance of prompt and non-prompt signatures discussed below.

The signal process $\gamma e^- \to e^+ \mu^- \mu^-$ proceeds dominantly via resonant production of an on-shell ALP in the subprocess $\gamma e^- \to a \mu^-$, followed by $a \to e^+ \mu^-$.
This leads to a characteristic resonance in the $e^+\mu^-$ invariant-mass distribution at $m_a$.

\begin{figure}[t]
    \centering
    \includegraphics[width=0.49\textwidth]{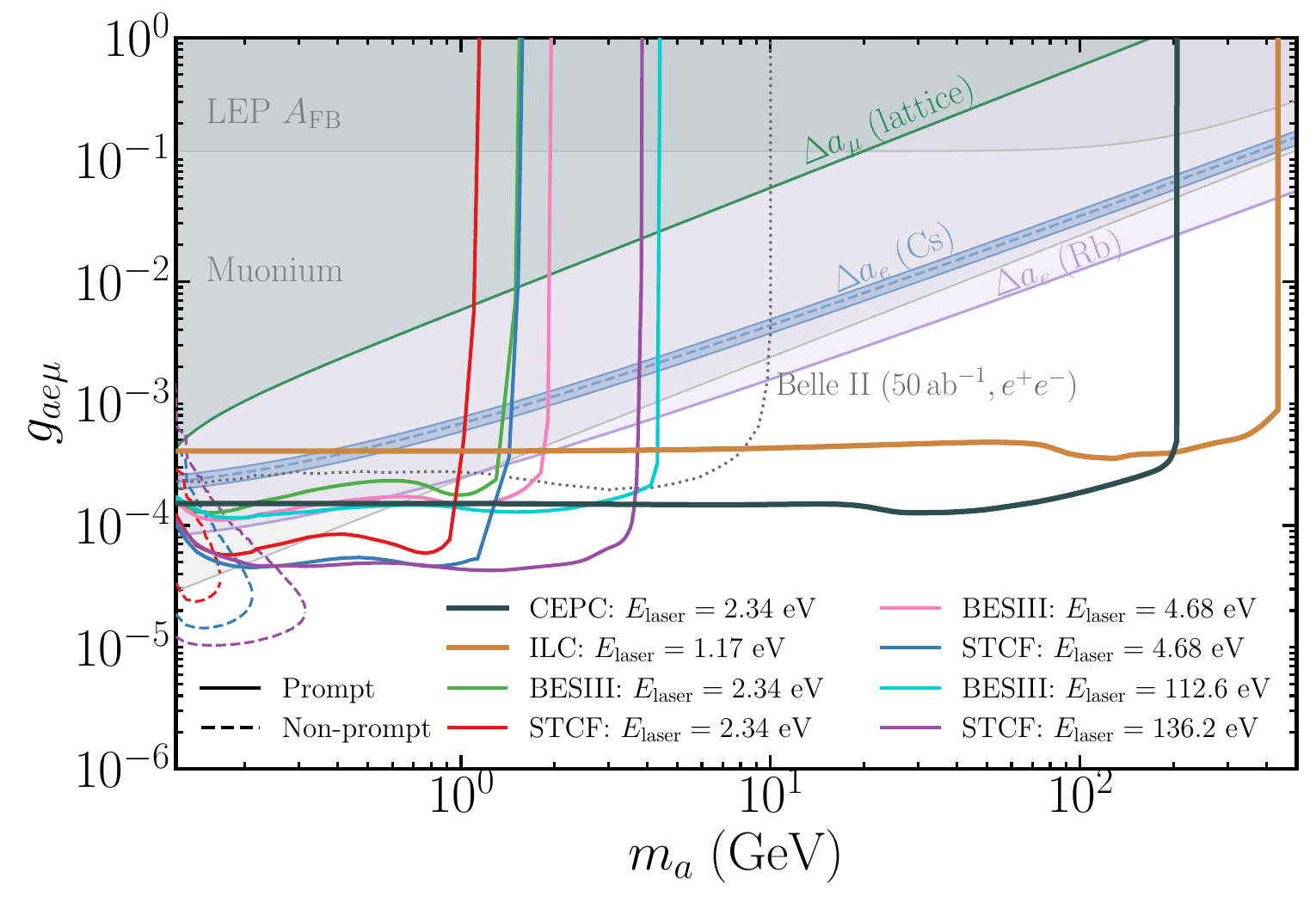}
    \caption{Projected $95\%$ C.L.~sensitivities to $g_{ae\mu}$ as functions of $m_a$ at BESIII, STCF, CEPC, and ILC. For comparison, the projected Belle~II sensitivity in the $e^+ e^-$ mode with an integrated luminosity of $50~\mathrm{ab}^{-1}$~\cite{Endo:2020mev} is also shown. Existing constraints from muonium--antimuonium oscillations~\cite{Willmann:1998gd}, leptonic $g-2$ (showing the $2\sigma$ constraint for $\Delta a_\mu$, $1\sigma$ favored band for $\Delta a_e$~(Cs), and the $3\sigma$ exclusion limit for $\Delta a_e$~(Rb))~\cite{Fan:2022eto,Muong-2:2025xyk,Aliberti:2025beg}, and LEP~\cite{ALEPH:2013dgf}, are included. The solid (dashed) curves correspond to the prompt (non-prompt) search strategies.}
    \label{fig:sensitivity}
\end{figure}

Existing constraints on $g_{ae\mu}$ arise from precision observables and collider measurements.
Muonium--antimuonium oscillations provide the dominant bounds in the mass range of interest, as the ALP mediates a tree-level four-fermion interaction, leading to a constraint that scales linearly with $m_a$~\cite{Willmann:1998gd}.
Loop-induced contributions to the anomalous magnetic moments of the muon and electron lead to highly complementary constraints.
These involve the recent experimental measurement of the muon $g-2$~\cite{Muong-2:2025xyk} and its SM prediction (where we explicitly adopt the lattice-QCD-based evaluation)~\cite{Aliberti:2025beg}, as well as the direct measurement of the electron $g-2$~\cite{Fan:2022eto} (whose theoretical prediction depends on the fine-structure constant $\alpha$ determined via Cesium~\cite{Parker:2018vye} and Rubidium~\cite{Morel:2020dww} atom interferometry).
A further constraint originates from collider bounds from LEP~\cite{ALEPH:2013dgf}.
We refer to Ref.~\cite{Endo:2020mev} for details.
These bounds are included in our numerical analysis and shown in Fig.~\ref{fig:sensitivity}, illustrating the complementarity between low-energy precision measurements and high-energy collider probes.

The flavor-violating structure of the coupling is essential for generating the same-sign dimuon final state already at leading order, while no SM process yields the same final state at the parton level.

\vspace{0.2cm}
\emph{\textbf{Electron--photon collider setup}---}
Electron--photon collisions are realized via laser Compton backscattering, where laser photons are scattered off the lepton beam to produce a high-energy photon beam.
We model the photon spectrum using the standard LCB formalism~\cite{Ginzburg:1981vm,Telnov:1989sd}, characterized by the parameter $\zeta = 4 E_e E_0 / m_e^2$, which determines the maximum photon energy fraction $x_{\rm max} = \zeta/(1+\zeta)$, where $E_e$ ($E_0$) denotes the electron (laser) beam energy.
For $\zeta \simeq 4.83$, the energy transfer is maximized.
We assume operation in the linear Compton scattering regime, where non-linear and pair-production effects are negligible.
The observable cross section is obtained by convoluting the partonic cross section with this photon distribution.

\begin{table}[t] 
    \centering
    \renewcommand{\arraystretch}{1.2} 
    \setlength{\tabcolsep}{1.5mm} 
    \begin{tabular}{l|cccc}
        \hline\hline
        Collider & $\sqrt{s}$~[GeV]& $E_e$~[GeV] & $E_0$~[eV] & $\mathcal{L}$~[ab$^{-1}$] \\
        \hline
        BESIII &5.6&2.8  & 2.34, 4.68, 112.6 & 0.02\\
        STCF &4.63& 2.315 & 2.34, 4.68, 136.2 & 1 \\
        CEPC & 240 & 120 & 2.34 & 20 \\
        ILC &500 & 250   & 1.17  & 4 \\
        \hline\hline
    \end{tabular}
    \caption{Benchmark parameters for BEPC-II (BESIII), STCF, CEPCss, and ILC in the $\gamma e$ collision mode.}
    \label{tab:collider_params}
\end{table}
We focus on four representative facilities: BEPC-II with the BESIII detector, STCF, CEPC, and ILC, whose benchmark parameters are summarized in Table~\ref{tab:collider_params}.
The chosen setups reflect mostly realistic operating conditions and well-motivated design benchmarks.

For BESIII, we assume a nominal collider energy $\sqrt{s}=5.6$~GeV (corresponding to a $e^+e^-$ center-of-mass (COM) energy of BEPC-II), together with an integrated luminosity of $20$~fb$^{-1}$ representative of current data-taking capabilities.
For STCF, we use $\sqrt{s}=4.63$~GeV, corresponding to a standard design operating point, with a projected integrated luminosity of $1$~ab$^{-1}$.
These facilities are therefore well suited to probe light ALPs in the sub-GeV to GeV mass range.

For CEPC, we take the baseline configuration of $\sqrt{s}=240$~GeV with an integrated luminosity of $20$~ab$^{-1}$.
For ILC, we adopt the baseline configuration at $\sqrt{s}=500$~GeV with an integrated luminosity of $4$~ab$^{-1}$.
The two relatively high-energy colliders can have sensitivity to significantly heavier ALPs.

For the corresponding laser configurations, the choices $E_0=2.34$ and $4.68$~eV for BESIII and STCF correspond to harmonics of solid-state lasers and are experimentally well motivated.
We also include higher-energy benchmarks ($\sim 100$~eV) to illustrate the kinematic reach in an idealized setup approaching $\zeta \simeq 4.83$.
For CEPC (ILC), a standard solid-state laser with $E_0 = 2.34$~eV ($E_0=1.17$~eV) is chosen, yielding $\zeta \simeq 4.3$ $(4.5)$, close to the optimal regime for energy transfer.

The four facilities span a wide range of energies and luminosities, providing complementary coverage of the ALP parameter space: BESIII and STCF probe light ALPs, while CEPC and ILC extend the reach to higher masses.

The sensitivity is driven by the interplay between the LCB photon spectrum and resonant ALP production.
The photon spectrum is enhanced near its kinematic endpoint, where the photon carries a large fraction of the beam energy, increasing the probability of accessing the on-shell production region in $\gamma e^- \to a \mu^-$.
Combined with the resonant enhancement, this leads to a significant increase in the signal rate compared to conventional $e^+e^-$ collisions, while maintaining a very clean experimental environment owing to the absence of irreducible SM backgrounds.

Details of the photon spectrum from LCB are provided in the Supplemental Materials.

\vspace{0.2cm}
\emph{\textbf{Signal and search strategies}---}
\begin{figure}[t]
\centering
\includegraphics[width=0.499\textwidth]{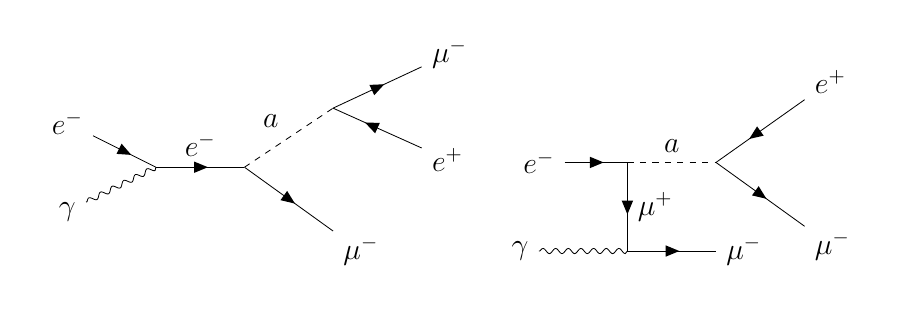}
\caption{Feynman diagrams for the signal process with on-shell ALP production.}
\label{fig:feynman}
\end{figure}
The signal process $\gamma e^- \to e^+\mu^-\mu^-$ proceeds dominantly via on-shell ALP production, as illustrated in Fig.~\ref{fig:feynman}, leading to a distinctive same-sign dimuon signature.

Signal events are generated using standard tools: the ALP model is implemented in \texttt{FeynRules}~\cite{Alloul:2013bka}, and events are simulated with \texttt{WHIZARD}~\cite{Kilian:2007gr} including the $\gamma e$ initial state.

The photon beam is modeled using the LCB spectrum, and the convolution with the partonic cross section is carried out on the event basis.
Detector acceptance is incorporated by requiring all visible leptons to satisfy basic kinematic criteria on the polar angle and momentum magnitude, consistent with the detector geometries of BESIII, STCF, CEPC, and ILC.

Owing to resonant ALP production, the invariant mass of one $e^+\mu^-$ pair reconstructs the resonance, providing a powerful handle for signal identification.

\begin{figure}[t]
    \centering
    \includegraphics[width=0.49\textwidth]{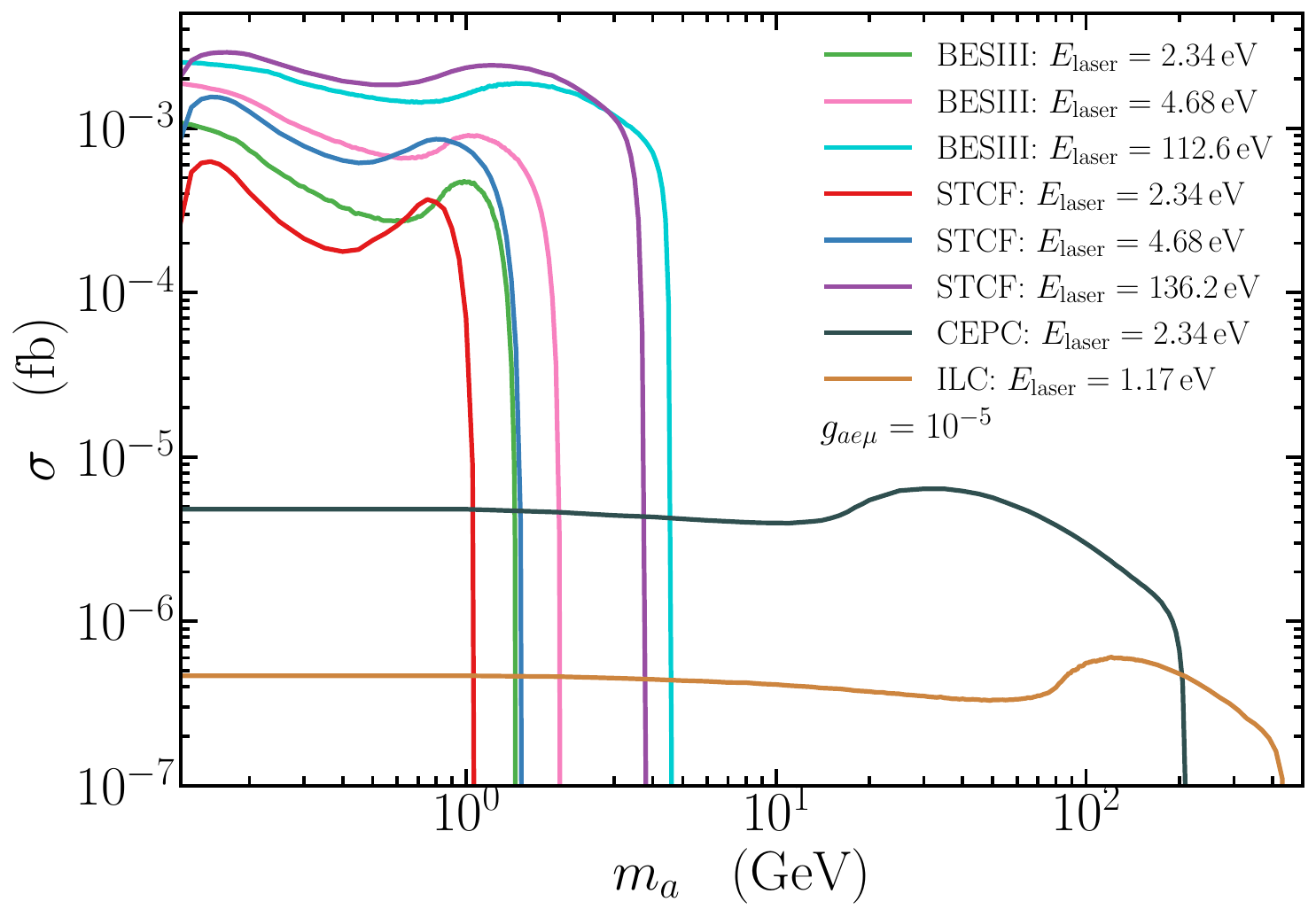}
    \caption{Cross sections of $\gamma e^- \to e^+ \mu^- \mu^-$ as functions of $m_a$ for BESIII, STCF, CEPC, and ILC, including detector acceptance and reconstruction efficiencies.}
    \label{fig:sigma_m}
\end{figure}

The signal cross sections for BESIII, STCF, CEPC, and ILC are shown in Fig.~\ref{fig:sigma_m}, including detector acceptance and reconstruction efficiencies.
For illustration, we fix $g_{ae\mu}=10^{-5}$.
The cross sections exhibit a non-monotonic dependence on $m_a$, reflecting the interplay between the LCB photon spectrum and the kinematic threshold for ALP production.
In the low-mass regime, the signal is enhanced by the peak of the photon spectrum.
As $m_a$ increases, larger photon energies are required and the cross section decreases, before rising again when the relevant kinematics approach the endpoint region where the photon spectrum is also enhanced.
Finally, as $m_a$ approaches the maximal accessible energy, phase-space suppression drives the cross section rapidly to zero.
The cross sections at CEPC and ILC are orders of magnitude below those at BESIII and STCF, mainly because at large COM energies the behavior of the cross sections is dominated by the parton-level cross sections, which scale as $1/s$~\cite{Huang:2025xvo}, instead of the LCB photon spectrum.

At the parton level, no SM process yields the same charge and flavor configuration, rendering the signal intrinsically clean.
After basic fiducial and kinematic selections, residual backgrounds arise only from detector-level effects, such as charge misidentification or secondary interactions, which are strongly suppressed in modern lepton detectors.
We note that, for the mass range considered here, $m_a \gtrsim 0.12~\mathrm{GeV}$, the angular separation between the $e^+$ and the $\mu^-$ originating from the ALP decay remains sufficiently large, and possible collimation effects are expected to be subleading, including at ILC.
In particular, the requirement of two same-sign muons together with an identified positron imposes a stringent constraint on charge reconstruction.
A quantitative estimate of such detector-induced backgrounds is provided in the Supplemental Materials, where we find that the residual background yields are at most $\mathcal{O}(1)$ event for BESIII and ILC, and remain moderate for STCF and CEPC depending on the laser configuration.
A narrow invariant-mass window around $m_a$ would further suppress the residual background significantly.

Residual backgrounds from unconverted $e^+e^-$ interactions are expected to be subleading compared to the charge-misidentification background discussed above and are therefore neglected in the present analysis.
Their quantitative impact depends sensitively on machine-specific beam-conversion and beam-removal efficiencies, and is thus beyond the scope of the present phenomenological study.

\begin{figure}[t]
    \centering
    \includegraphics[width=0.499\textwidth]{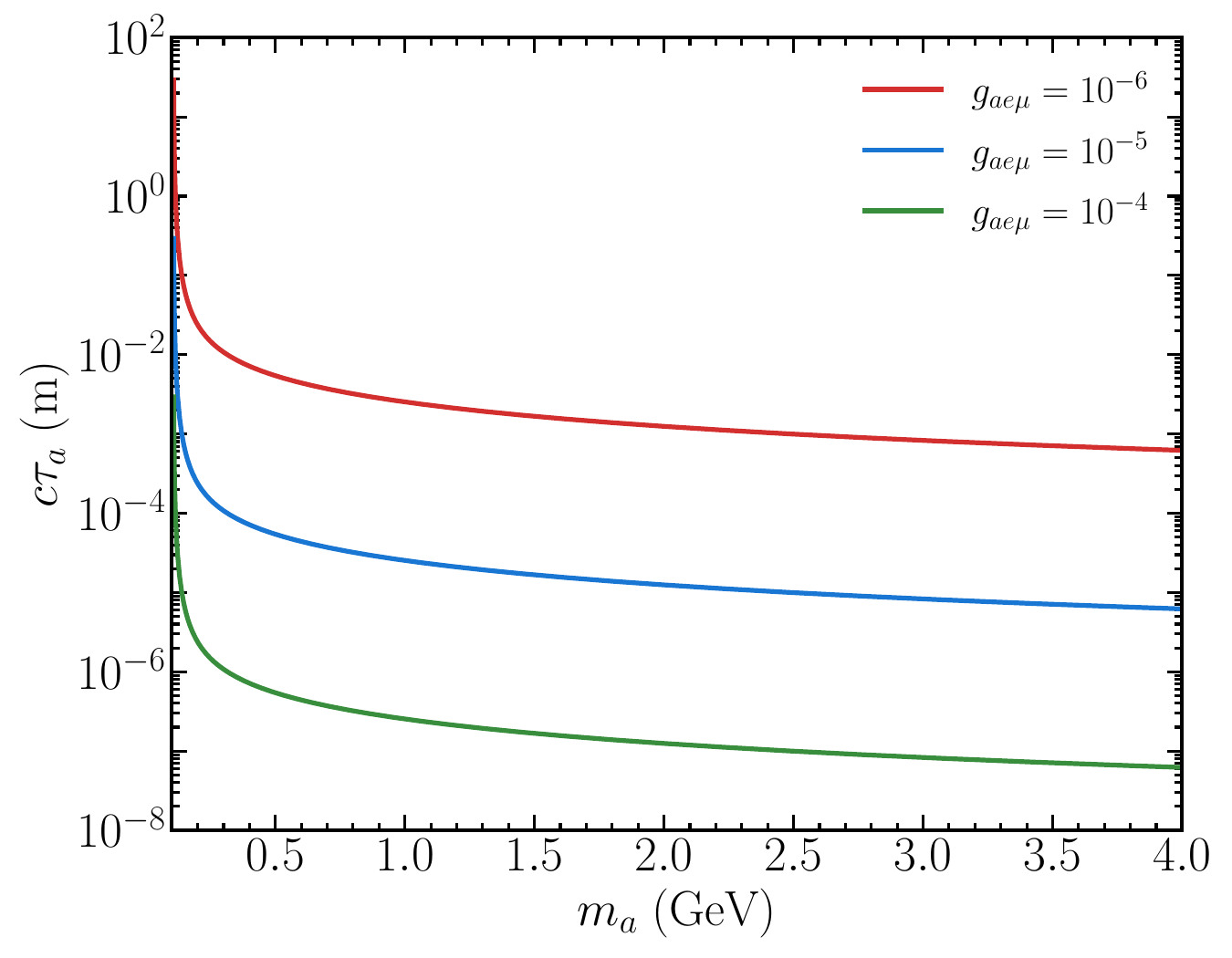}
    \caption{Proper decay length of the ALP as a function of its mass for representative values of $g_{ae\mu}$, illustrating the transition between prompt and non-prompt regimes relevant for different collider setups.}
    \label{fig:ctau}
\end{figure}

Depending on the ALP lifetime, two complementary search strategies can be employed.
For sufficiently large couplings, the ALP decays promptly, yielding a clean $e^+\mu^-\mu^-$ signature, as described above.
For smaller couplings, it can be long-lived and produce a non-prompt decay with a displaced vertex (DV) signature inside the detector.
As shown in Fig.~\ref{fig:ctau}, the proper decay length varies significantly across the parameter space, giving rise to distinct prompt and non-prompt regimes.

We find that the non-prompt strategy is particularly relevant for STCF, while for BESIII, CEPC, and ILC it does not have sensitivities, except for the case of an idealized $\sim 100$~eV laser-beam setup at BESIII.
Accordingly, our numerical analysis focuses on STCF for the non-prompt search.
For even longer lifetimes the ALP could appear as missing energy at the detector; however, our studies indicate that searches based on missing-energy signatures are ineffective for all four colliders.

This pattern reflects the interplay between the ALP lifetime, the boost in the asymmetric $\gamma e^-$ setup, and the signal rate.
For light ALPs at BESIII and STCF, the boost is moderate and the decay length is mainly controlled by the coupling strength, so both prompt and displaced signatures can be relevant. 
At ILC and CEPC, however, the signal cross section after kinematic selections is significantly smaller, making the experiment sensitive primarily to larger values of $g_{ae\mu}$.
In this regime, the ALP lifetime is short, so it typically decays promptly despite its large boost.
As a result, the parameter region with observable displaced signatures is strongly reduced at high energies.

Backgrounds for the non-prompt search are expected to be strongly suppressed, as no SM process can produce a DV with the same charge and flavor configuration.
Detector-induced backgrounds, such as fake vertices or secondary interactions, are expected to be negligible after standard vertex-quality requirements.
Thus, for the STCF setup, we treat the non-prompt search as effectively background-free in a conservative approximation.

The corresponding sensitivities are derived by incorporating detector acceptance, reconstruction efficiencies, and the residual backgrounds discussed above.
Details of the simulation setup, event selection (including kinematic acceptance), fiducial-volume definitions for prompt and displaced decays, and background estimation are given in the Supplemental Materials.

\vspace{0.2cm}
\emph{\textbf{Sensitivity reach}---}
Fig.~\ref{fig:sensitivity} shows the projected $95\%$ C.L.~sensitivities to $g_{ae\mu}$ as functions of $m_a$.
For comparison, we also include the projected sensitivity from Belle~II in the $e^+ e^-$ mode with an integrated luminosity of 50~ab$^{-1}$~\cite{Endo:2020mev}.
We find that the $\gamma e$ collider setups, despite operating at much lower integrated luminosities, can probe values of $g_{ae\mu}$ that are more than one order of magnitude smaller over a wide mass range, demonstrating the clear advantage of the $\gamma e$ collision mode.
Existing constraints from muonium--antimuonium oscillations~\cite{Willmann:1998gd}, LEP measurements~\cite{ALEPH:2013dgf}, and leptonic $g-2$~\cite{Fan:2022eto,Muong-2:2025xyk,Aliberti:2025beg,Parker:2018vye,Morel:2020dww} are also displayed.
Specifically, these bounds have been systematically re-evaluated using the most recent experimental data and SM predictions.
The analytical expressions for the leptonic $g-2$ evaluation are detailed in the Supplemental Materials, while the constraints for muonium--antimuonium oscillations and LEP measurements are computed using the formulas given in Ref.~\cite{Endo:2020mev}.
It is worth noting that the narrow blue band in Fig.~\ref{fig:sensitivity} specifically denotes the $1\sigma$ favored regions that could accommodate the observed experimental anomalies for the Cs-based $\Delta a_e$.
In contrast, the Rb-based measurement yields a positive $\Delta a_e$ shift, which is not allowed by the negative contribution from the cLFV ALP; thus, it is incorporated as a $3\sigma$ upper exclusion limit (purple region).
Further, since the latest lattice-based calculation of the muon $g-2$ is in agreement with the most recent experimental measurement, we show the $2\sigma$ upper bound on $g_{a e \mu}$ in the green-filled region.

Sensitivities are derived using a counting-based criterion, requiring three signal events in the near-zero-background regime and a significance of $Z=2$ otherwise, based on the background estimates provided in the Supplemental Materials.
Here, $Z=2$ is used as the sensitivity criterion, corresponding to an approximate $2\sigma$ statistical significance under the Gaussian approximation.
The analytic expression of $Z$ is given in the Supplemental Materials.
The projected sensitivities assume dedicated running in the $\gamma e$ collision mode, and therefore scale with the achievable integrated luminosity in this configuration.

STCF, CEPC, and ILC probe $g_{ae\mu}$ one to two orders of magnitude below current bounds over a wide mass range, covering previously unexplored regions of parameter space, reaching up to $\mathcal{O}(1)$, $\mathcal{O}(100)$, and $\mathcal{O}(100)$~GeV, respectively, while BESIII probes a more limited region.

Detector-induced backgrounds reduce the sensitivity of the prompt search at STCF and CEPC, which dominates at relatively large couplings.
In contrast, the non-prompt search at STCF is more sensitive in the small-coupling and low-mass region.
The two strategies therefore provide complementary coverage of the parameter space.
Across different facilities, the broad range of COM energies extends the ALP mass reach.

\vspace{0.2cm}
\emph{\textbf{Conclusions and outlook}---}
We have identified a previously unexplored and intrinsically clean same-sign dimuon signature in electron--photon collisions, $\gamma e^- \to e^+\mu^-\mu^-$, mediated by an axionlike particle with $e$--$\mu$ flavor-violating couplings.
The absence of irreducible SM backgrounds and the resonant production of the mediator render this channel exceptionally sensitive, with only small residual backgrounds arising from detector effects.
Such backgrounds can affect the prompt sensitivity at STCF and CEPC.
For STCF, a non-prompt (displaced) search provides additional sensitivity in part of the parameter space, leading to complementary coverage.

We have shown that electron--photon collisions at STCF, CEPC, and ILC can probe $g_{ae\mu}$ values one to two orders of magnitude below existing bounds, covering previously unexplored regions of parameter space.
This sensitivity improves upon conventional $e^+e^-$ searches such as Belle~II by up to one order of magnitude, despite their much larger integrated luminosities, highlighting the unique advantage of the $\gamma e$ collision mode.

More broadly, this work demonstrates that electron--photon collisions provide a qualitatively new and uniquely powerful framework for probing charged lepton flavor violation.
This strategy exploits the interplay between resonant production and background-free or low-background final states, and can be extended to other flavor structures and mediator types, opening a new direction for flavor-violating searches at future colliders.

\vspace{0.2cm}
\emph{\textbf{Acknowledgments}---}
We would like to thank Xiaorong Zhou for useful discussions on BESIII and STCF. This work was supported by the National Natural Science Foundation of China under grants No.~12475106 and 12505120, and the Fundamental Research Funds for the Central Universities under Grant No.~JZ2025HGTG0252.

\bibliography{refs}

\newpage

\widetext
\begin{center}
   \textbf{\large SUPPLEMENTAL MATERIAL \\[.2cm] ``Same-sign dimuon probe of charged lepton flavor violation at electron–photon colliders''}\\[.2cm]
  \vspace{0.05in}
  {Zhong Zhang, Yu Zhang, and Zeren Simon Wang}
\end{center}
\setcounter{equation}{0}
\setcounter{figure}{0}
\setcounter{table}{0}
\setcounter{page}{1}
\setcounter{section}{0}
\makeatletter
\renewcommand{\thesection}{S-\Roman{section}}
\renewcommand{\theequation}{S-\arabic{equation}}
\renewcommand{\thefigure}{S-\arabic{figure}}
\renewcommand{\thetable}{S-\arabic{table}}
\renewcommand{\bibnumfmt}[1]{[S#1]}
\renewcommand{\citenumfont}[1]{#1}

\section{Photon spectrum from laser Compton backscattering}
In the $\gamma e$ collision mode, high-energy photons are generated via Compton backscattering of laser photons off the lepton beam.
Denoting the lepton-beam energy by $E_e$ and the laser-beam energy by $E_0$, the energy of the backscattered photon is $E_\gamma = x E_e$, where $x$ follows the distribution $f_{\gamma/e}(x)$.

The observable cross section is obtained by convoluting the partonic cross section with the photon spectrum,
\begin{equation}
    \frac{d\sigma}{d\cos\theta} = \int_{0}^{x_{\rm max}} dx \, f_{\gamma/e}(x) \, \frac{d\hat{\sigma}}{d\cos\theta},
\end{equation}
where $\theta$ is the scattering angle in the partonic COM frame.
The unpolarized photon spectrum $f_{\gamma/e}(x)$ is given by~\cite{Ginzburg:1981vm,Telnov:1989sd}
\begin{equation}
    f_{\gamma/e}(x) = \frac{1}{g(\zeta)} \left[ 1 - x + \frac{1}{1 - x} - \frac{4x}{\zeta (1 - x)} + \frac{4x^2}{\zeta^2 (1 - x)^2} \right],
\end{equation}
with
\begin{equation}
    g(\zeta) = \left( 1 - \frac{4}{\zeta} - \frac{8}{\zeta^2} \right) \ln(\zeta + 1) + \frac{1}{2} + \frac{8}{\zeta} - \frac{1}{2(\zeta + 1)^2}.
\end{equation}
The parameter $\zeta = 4E_e E_0 / m_e^2$ controls the kinematics, with the maximal photon energy fraction given by $x_{\rm max} = \zeta/(1+\zeta)$.

\section{Monte Carlo simulation setup}
We implement the ALP effective interaction in \texttt{FeynRules}~\cite{Alloul:2013bka} and export the corresponding UFO model~\cite{Degrande:2011ua}.
Signal events for $\gamma e^- \to e^+ \mu^- \mu^-$ are generated using \texttt{WHIZARD}~\cite{Kilian:2007gr} at the parton level including detector acceptances.
The LCB photon spectrum is incorporated by convoluting the partonic cross section with the analytical photon distribution described above.
Reconstruction efficiencies are then applied at the analysis level.

For BESIII, we require $|\cos\theta|<0.93$ for both electrons and muons, and apply a flat identification efficiency of $90\%$ for muons with $p>0.5$~GeV~\cite{BESIII:2009fln}.
For STCF, we assume $|\cos\theta|<0.93$ and apply momentum-dependent tracking efficiencies following Ref.~\cite{Achasov:2023gey}.
For muons, the tracking efficiency is taken to be zero for $p<0.4$~GeV, to increase linearly to $70\%$ at $0.5$~GeV and to $95\%$ at $0.7$~GeV, and to remain at $95\%$ for $p>0.7$~GeV.
For electrons, we require $p_T>0.1$~GeV, with the efficiency rising linearly from $90\%$ to $99\%$ at $0.3$~GeV and remaining at $99\%$ thereafter.

For ILC, we require $|\cos\theta|<0.95$ and $|p|>0.1$~GeV, with flat efficiencies $\epsilon_e=0.97$ and $\epsilon_\mu=0.98$, consistent with the expected performance of precision tracking detectors~\cite{ILCInternationalDevelopmentTeam:2022izu}.
For CEPC, we assume $|\cos\theta|<0.99$ with a tracking efficiency of $0.99$, and $|p|>5$~GeV with a lepton identification efficiency of $0.99$, referring to Ref.~\cite{CEPCStudyGroup:2025kmw}.

\section{Fiducial volumes and decay treatment}
The treatment of ALP decays depends on its boosted transverse decay length $L^a_T=\beta_T\gamma c\tau_a$ and on detector resolution and geometry, leading to distinct prompt and non-prompt decay regimes.

For the prompt search, we require the ALP to decay within a region experimentally indistinguishable from the interaction point.
This is implemented by introducing an effective prompt-decay scale $R_{\rm res}$, characterizing the typical spatial/vertex resolution of the tracking system.
Each event is weighted by
\begin{equation}
    P_{\rm prompt} = 1 - \exp\left(-\frac{R_{\rm res}}{L^a_T}\right).
\end{equation}
For BESIII, we take $R_{\rm res} = 0.13~\mathrm{mm}$~\cite{BESIII:2009fln}.
For STCF, we assume conservatively $R_{\rm res} = 0.1~\mathrm{mm}$, consistent with the projected detector performance~\cite{Achasov:2023gey}.
Decays within this region are effectively indistinguishable from prompt leptons at the analysis level.
The scale $R_{\rm res}$ should be interpreted as an effective separation scale rather than a strict detector threshold.

For ILC and CEPC, we do not explicitly impose a prompt-decay requirement, as the ALP typically decays promptly in the accessible parameter space.
A realistic separation between prompt and displaced decays would require detailed detector modeling, which is beyond the scope of the present analysis.

Complementarily, we consider a non-prompt (displaced) decay strategy by requiring the decay to occur outside the prompt region and within a fiducial tracking volume.
The corresponding probability is
\begin{equation}
P_{\text{NP}} = \frac{1}{L^a_T}\int_{R_{\text{min}}}^{L_T} dr \, e^{-r/L^a_T} \, \epsilon_{\text{vtx}}(r),
\end{equation}
where $\epsilon_{\text{vtx}}(r)=1-r/R_{\rm max}$ provides a simple linear parameterization of the vertex-reconstruction efficiency as a function of the decay radius~\cite{Dib:2019tuj,Cheung:2021mol}.
Here,
\begin{equation}
    L_T \equiv \min\!\left(\max(R_{\text{min}},\,|L_d\tan\theta|),\,R_{\text{max}}\right),
\end{equation}
with $L_d$ the half length of the main drift chamber (MDC) and $\theta$ the ALP polar angle.

A meaningful non-prompt sensitivity at BESIII would require an idealized high-energy laser configuration (e.g.\ $E_0 \simeq 112.6$~eV), which we do not consider realistic.
We therefore focus on STCF for this strategy, adopting
\begin{equation}
L_d=1400~\text{mm}, \qquad R_{\text{min}}=0.1~\text{mm},\qquad R_{\text{max}}=850~\text{mm},
\end{equation}
following Ref.~\cite{Achasov:2023gey}.
This definition captures decays that occur outside the prompt region and can lead to observable deviations from prompt-like kinematics, and are contained within the tracking volume defined by the outer radius of the MDC ($R_{\rm max}$) and its half-length $L_d$.

\section{Estimate of detector-induced backgrounds for the prompt searches}
As a quantitative assessment of detector-induced backgrounds to the prompt searches, we consider the SM process $\gamma e^- \to e^- \mu^+ \mu^-$, which constitutes the dominant reducible background with the closest kinematic topology to the signal.

After imposing the same kinematic selections as in the signal analysis, this process can mimic the signal only if both the final-state $e^-$ and $\mu^+$ are reconstructed with wrong charges.
In realistic magnetic tracking systems, charge misidentification is expected to be rare.
We conservatively assume a benchmark charge-misidentification probability of $\mathcal{O}(10^{-3})$ per lepton~\cite{BESIII:2009fln,Achasov:2023gey,ILCInternationalDevelopmentTeam:2022izu}, corresponding to an overall double-charge-misidentification suppression factor of $\mathcal{O}(10^{-6})$.
This choice is consistent with the expected performance of modern tracking detectors and is therefore conservative.
A more precise determination of this rate would require a dedicated detector simulation and reconstruction study, which is beyond the scope of the present work.

\begin{table}[h]
\centering
\begin{tabular}{l|ccc|c}
\hline\hline
 & \multicolumn{3}{c}{Background events} &  \\
Collider & low $E_0$ & medium $E_0$ & high $E_0$ & $\mathcal{L}$~[ab$^{-1}$] \\
\hline
BEPC-II (BESIII) & 0.9 & 2.8 & 8.0 & 0.02 \\
STCF   & 36.8 & 77.4 & 476.1 & 1 \\
CEPC & 100.1 & -- & -- & 20 \\
ILC    & 1.0 & -- & -- &   4 \\
\hline\hline
\end{tabular}
\caption{Estimated residual background yields after all selections for different collider and laser configurations. The results correspond to the integrated luminosities assumed in the main text.}
\label{table:bgd_N}
\end{table}
With this conservative assumption, the resulting residual background yields after all selections are summarized in Table~\ref{table:bgd_N}.

The prompt searches remain effectively in the near-zero-background regime for the ILC setup and for the lower-laser-energy BESIII configurations, while nonzero residual backgrounds are expected for STCF and CEPC, and for the highest-laser-energy BESIII benchmark.
In the main text, we account for this effect using a counting-based sensitivity criterion, requiring three signal events in the near-zero-background regime and $Z=2$ otherwise.
The signal significance $Z$ is defined as
\begin{equation}
Z = \sqrt{2 \cdot \left((s + b) \cdot \ln(1 + s/b) - s\right)} ,
\end{equation}
where $s$ and $b$ label the expected number of signal and background events, respectively.

These estimates are conservative and likely overestimate the true background rates.
In a realistic experimental analysis, the residual background could be further reduced by optimized event selections, such as tighter kinematic requirements, including an invariant-mass window around the reconstructed  $e^+ \mu^-$ system, or improved reconstruction strategies.
Since the quantitative impact of such optimizations depends on detector-specific effects, including momentum- and angle-dependent charge-misidentification rates as well as the mass resolution, we do not attempt to implement them here and instead apply a conservative approach.

\section{Analytical Formulas for Lepton $(g-2)$ Constraints}
\label{app:g-2_formulas}
The one-loop contributions of the cLFV ALP to the anomalous magnetic moments of the electron and muon are evaluated using the expressions given below.

The ALP contribution to the electron $g-2$ ($\Delta a_e$) is predominantly driven by the intermediate muon loop, yielding:
\begin{equation}
\Delta a_e = -\frac{1}{16\pi^2} \frac{m_e}{m_\mu} |g_{ae\mu}|^2 \left[ \frac{2x^2 \ln x}{(x-1)^3} + \frac{1-3x}{(x-1)^2} \right],
\end{equation}
where $x = m_a^2 / m_\mu^2$.

Similarly, the one-loop ALP contribution to $\Delta a_\mu$ is expressed as:
\begin{equation}
\Delta a_\mu = \frac{1}{16\pi^2} |g_{ae\mu}|^2 \left[ 2x^2 \ln\left(\frac{x}{x-1}\right) - 2x - 1 \right].
\end{equation}

The direct measurement of the electron $g-2$~\cite{Fan:2022eto} (whose theoretical prediction depends on the fine-structure constant $\alpha$ determined via Cesium~\cite{Parker:2018vye} and Rubidium~\cite{Morel:2020dww} atom interferometry) yields~\cite{Aliberti:2025beg}
\begin{align}
a_e^\text{exp} - a_e^\text{SM}\left[\alpha(\text{Cs})\right] &= -1.00(26) \times 10^{-12}, \nonumber \\
a_e^\text{exp} - a_e^\text{SM}\left[\alpha(\text{Rb})\right] &= +0.35(16) \times 10^{-12}.
\end{align}
For the muon $g-2$, we use the recent experimental measurement~\cite{Muong-2:2025xyk} and its SM prediction based on lattice QCD calculations~\cite{Aliberti:2025beg}, defining
\begin{align}
\Delta a_\mu \equiv a_\mu^\text{exp} - a_\mu^\text{SM} = 38(63) \times 10^{-11}.
\end{align}

\section{Constraints from $\mu\to e\gamma$ in the presence of diagonal ALP couplings}

In the main text we focus on the simplified scenario in which the cLFV coupling
$g_{ae\mu}$ is the only non-vanishing ALP interaction with charged leptons.
More generally, cLFC couplings to electrons and muons may also
be present. In such a case, the radiative decay $\mu\to e\gamma$ is induced at
one loop and can provide important complementary constraints.

Following Ref.~\cite{Calibbi:2024rcm}, the decay width can be written as
\begin{equation}
\Gamma(\mu \to e\gamma) = \frac{\alpha m_\mu}{2 (32\pi^2)^2} \left| g_{ae\mu} \right|^2 \left| g_{\mu\mu} \, g_1(x_\mu) + g_{ee} \, g_2(x_\mu) \right|^2,
\end{equation}
where $x_\mu = m_a^2 / m_\mu^2$, and the loop functions are given by:
\begin{align}
g_1(x) &= \frac{(x - 3)x^2 \log x}{x - 1} - 2x + 1 - 2\sqrt{x - 4}\,x^{3/2} \log\!\left(\frac{\sqrt{x - 4} + \sqrt{x}}{2}\right), \\
g_2(x) &= 1 - 2x + 2(x - 1)x \log\!\left(\frac{x}{x - 1}\right).
\end{align}
Using the latest MEG II limit~\cite{MEGII:2025gzr},

\begin{equation}
\mathrm{BR}(\mu\to e\gamma)
=
\frac{\Gamma(\mu\to e\gamma)}
{\Gamma_\mu}
<
1.5\times10^{-13},
\end{equation}

one obtains

\begin{equation}
|g_{e\mu}|
\,
\left|
g_{\mu\mu}g_1(x_\mu)
+
g_{ee}g_2(x_\mu)
\right|
<
3.41\times10^{-12}.
\end{equation}

We note that cLFC ALP couplings are also constrained by a variety of direct searches and precision observables.
In particular, existing limits on the electron coupling can be derived from $e^+e^-$ collider searches and precision measurements, while constraints on the muon coupling arise from anomalous magnetic moment measurements and other low-energy observables.
A comprehensive review of these bounds can be found in Ref.~\cite{Bauer:2017ris}.
The purpose of the present section is not to perform a global analysis of flavor-conserving ALP interactions, but rather to quantify the additional restrictions imposed by the cLFV decay $\mu\to e\gamma$ when both cLFV and cLFC couplings are simultaneously present.

\begin{figure}[t]
    \centering
    \includegraphics[width=0.49\textwidth]{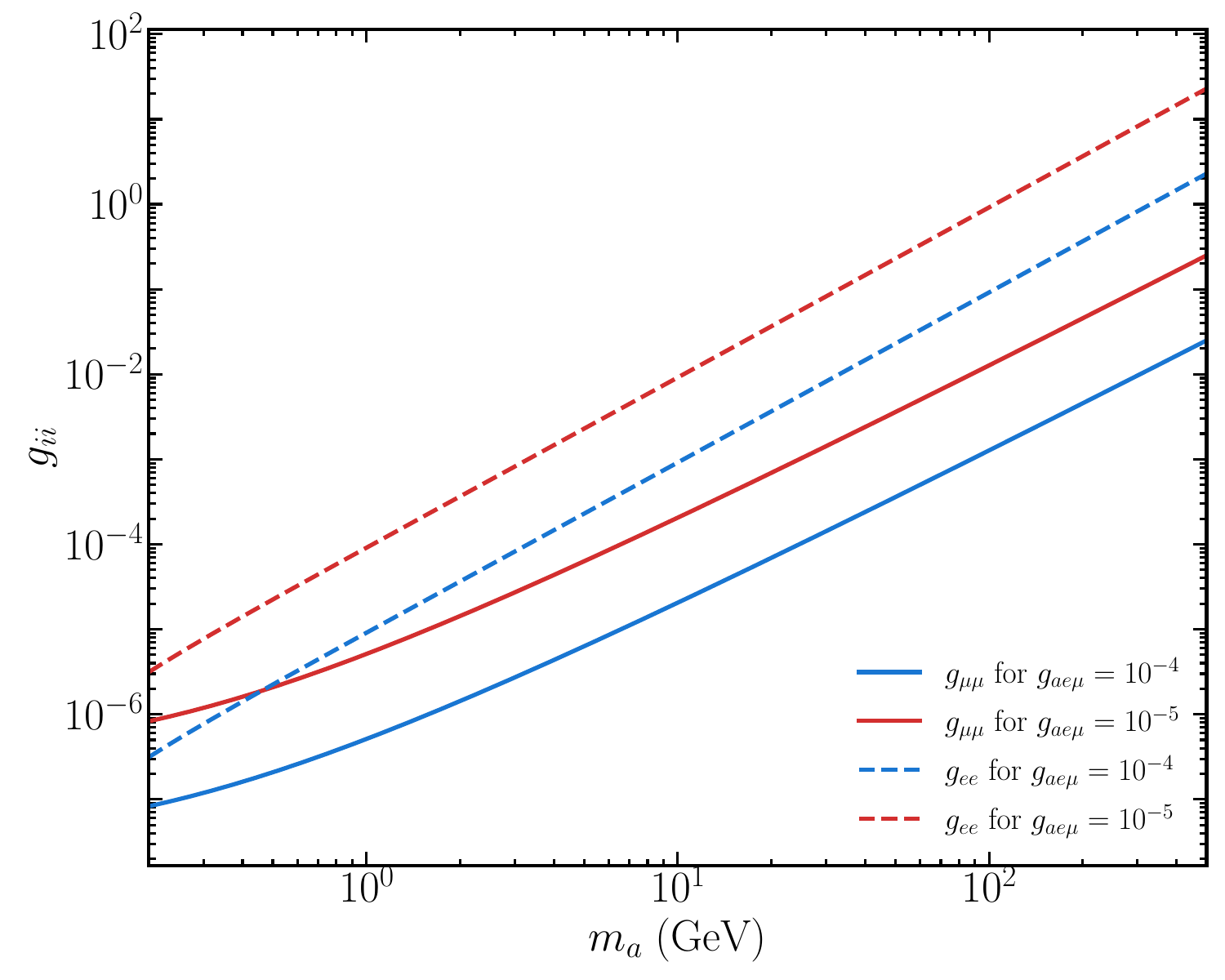}
    \caption{Constraints on the flavor-conserving ALP couplings from the MEG II bound on $\mu\to e\gamma$. Left: upper limits on $|g_{\mu\mu}|$ assuming $g_{ee}=0$. Right: upper limits on $|g_{ee}|$ assuming $g_{\mu\mu}=0$. The blue and red curves correspond to $|g_{ae\mu}|=10^{-4}$ and $10^{-5}$, respectively.}
    \label{fig:megii}
\end{figure}

To illustrate the implications of this bound, we consider two benchmark
scenarios:

\begin{itemize}
\item $g_{ee}=0$ and $g_{\mu\mu}\neq0$;
\item $g_{\mu\mu}=0$ and $g_{ee}\neq0$.
\end{itemize}

The resulting upper limits on the flavor-conserving couplings are shown in Fig.~\ref{fig:megii} for representative values $g_{ae\mu}=10^{-4}$ and $10^{-5}$.

\end{document}